%% file: ht2017.tex
\documentclass[sigconf]{acmart}

\usepackage{booktabs} % For formal tables
\usepackage{multirow, array}
\usepackage{pbox}
\usepackage{amsmath}
\usepackage{flushend}

\newtheorem{problem}{Problem}

\copyrightyear{2017}
\acmYear{2017}
\setcopyright{acmcopyright}
\acmConference{HT '17}{July 04-07, 2017}{Prague, Czech Republic}
\acmPrice{15.00}
\acmDOI{http://dx.doi.org/10.1145/3078714.3078735}
\acmISBN{978-1-4503-4708-2/17/07}

\begin{document}
\title[Detection of Trending Topic Communities]{
Detection of Trending Topic Communities:\\
Bridging Content Creators and Distributors}

\author{Lorena Recalde, David F. Nettleton,\newline Ricardo Baeza-Yates}

\affiliation{%
  \institution{Web Research Group, DTIC, Universitat Pompeu Fabra}
  \streetaddress{Roc Boronat, 138}
  \city{Barcelona} 
  \state{Spain} 
  \postcode{08018}
}
\email{{lorena.recalde, david.nettleton, ricardo.baeza}@upf.edu}

\author{Ludovico Boratto}
\orcid{0000-0002-6053-3015}
\affiliation{%
  \institution{Digital Humanities}
  \institution{EURECAT}
  \streetaddress{Av. Diagonal 177}
  \city{Barcelona} 
  \state{Spain} 
  \postcode{08018}
}
\email{ludovico.boratto@acm.org}

\renewcommand{\shortauthors}{L. Recalde et al.}

\begin{abstract}
The rise of a trending topic on Twitter or Facebook leads to the temporal emergence of a set of users currently interested in that topic. Given the temporary nature of the links between these users, being able to dynamically identify communities of users related to this trending topic would allow for a rapid spread of information. Indeed, individual users inside a community might receive recommendations of content generated by the other users, or the community as a whole could receive group recommendations, with new content related to that trending topic. In this paper, we tackle this challenge, by identifying coherent topic-dependent user groups, linking those who generate the content ({\em creators}) and those who spread this content, {\em e.g.}, by retweeting/reposting it ({\em distributors}). This is a novel problem on group-to-group interactions in the context of recommender systems. Analysis on real-world Twitter data compare our proposal with a baseline approach that considers the retweeting activity, and validate it with standard metrics. Results show the effectiveness of our approach to identify communities interested in a topic where each includes content creators and content distributors, facilitating users' interactions and the spread of new information.
\end{abstract}




\maketitle

\input{body-ht2017}
\bibliographystyle{ACM-Reference-Format}
\bibliography{references} 
%\balance
\end{document}

%% file: body-ht2017.tex
\section{Introduction}

Once we belong to an online social network (OSN) we can share content, add people to our network, access interesting information streams created by relevant users, and express our likes and comments about items shared by other users. Personalization is a key feature in OSNs because not all the content generated by our connections may be of our interest, regardless of its quality. Likewise, not all of our connections generate content that we might consider adequate, even if it fits into our topics of interest.

In order to enhance personalization, social recommender systems as part of OSNs are in charge of filtering content streams based on each user's interests model, their trusted social connections activity, and content authority. To do this, one way of finding relevant items to recommend to a user would be to discover their meaningful connections. For instance, the degree of significance could be measured in terms of the impact of the resources the user shares and the links the user has with those inside a topic-dependent community. 

When a word, a phrase, or a hashtag is used with a high frequency, it is said to be associated to a {\em trending topic}. With the rise of a trending topic, a set of users interested in it also emerges. However, multiple points of view might be associated to it ({\em e.g.}, the {\tt \#donaldtrump} hashtag, related to the recently-elected US president, is used by people with opposing political views). Being able to manage these users and detect communities associated to a given trending topic is a problem of central interest in social recommender systems. Indeed, having a community of users who are linked and have the same interests would allow a system to generate suggestions at multiple granularities, {\em i.e.}, ($i$) for individual users, by providing recommendations of content related to the trending topic and generated by the other users in the community (thus allowing a quick and effective spread of information); or ($ii$) for the community as a whole, by providing group recommendations with new content related to the trending topic. At the same time, the problem is challenging, since trending topics are characterized by their temporary nature and evolve quickly; therefore, an approach that detects communities in this context should run quickly ({\em i.e.}, have a fast processing time), in order to dynamically adapt to the evolution of the trending topic (for example, by considering new users interested in it).

In order to tackle the problem of detecting communities related to a trending topic, in this paper we focus on Twitter, the widely-known microblogging platform. The activity of Twitter is depicted by tweets, retweets, replies, likes and shares, and its structure is defined by \textit{follower} and \textit{followee} unidirectional relationships.
A key characteristic of Twitter, and of our approach, in order to enable the desired spread of information, is following and being followed by other users. Follower users are interested in tracking down significant users to follow, whereas the followed (leader) users wish to accumulate a lot of followers. However, to create significant content and be a topic influential user it is necessary to obtain interesting, trendy, and relevant information to generate a tweet. One way of doing this is to form a ``collusion" with other content creators or influencers in the domain. As a result, the influential group is able to share and filter key news before they become widely known, and then potentiate its diffusion through the group of users interested in that topic (who may have the role of distributors or consumers of the given topic).

Accordingly, we present a method to identify groups of topic-dependent ``content creators" (\textit{CCs}) in Twitter. Another key element of our proposal is the identification of their matching spreader groups or topic-dependent ``content distributors" (\textit{CDs}). After the identification of these two categories of users, both $CCs$ and $CDs$ are linked by our approach in a unique community, which represents the user base for the different forms of recommendation previously mentioned. 

In summary, given this real-world application scenario, our objective is to detect communities of users who ($i$) are associated to a given trending topic, ($ii$) are interested in the same content, ($iii$) are linked among themselves ({\em i.e.}, they follow each other), and ($iv$) can be either identified as content creators or content distributors. 

Formally, the problem statement is the following:
\begin{problem}
Let $H$ be the set of trending topics at a given time. For each topic $h \in H$, let $T_h$ be the set of tweets that contain $h$ ({\em i.e.}, those associated to the trending topic), and $U_h$ be the set of users who posted a tweet that belongs to $T_h$. The first goal is to identify a set of {\em content creators} $CCs \subseteq U_h$, who generated tweets that have been retweeted multiple times. The second goal is the identification of a set of {\em content distributors} $CDs \subseteq U_h$, who retweeted content generated by a $CC$. The final goal is building a graph $G$ that contains the $CCs$ and $CDs$ as vertices, connected by edges that represent the ``following'' and ``who-retweeted-who'' relationships, which will allow us to detect communities that contain both $CCs$ and $CDs$.
\end{problem}

To the best of our knowledge, our work represents the first attempt to detect several communities interested in a given topic, where each community integrates both a content creator group and the corresponding distributor group. The proposed method would improve the interaction and communication among the members of the community, and may be used to generate more personalized recommendations based on the structure of the topic-based community and levels of social influence. To summarize, our contributions are:

\begin{itemize}
\item We define a social model that detects topic-dependent content creator and content distributor groups on Twitter;
\item The model can be embedded in an individual or group recommender system to suggest social entities;
\item We validate our proposal on a real-world dataset extracted from Twitter, by employing standard metrics and by comparing it with a baseline approach that only requires the retweeting activity. 
\end{itemize}

The remainder of the paper is organized as follows: Section~\ref{related} summarizes the context of the present work and the related state of the art; Section~\ref{approach} describes our approach; in Section~\ref{framework} we present the analytical framework built to validate our proposal and the obtained results; finally, in Section~\ref{conclusions}, we conclude and propose future work.

\section{Background and Related Work}\label{related}

The Social Web has shown to be one of the richest sources for mining people's interests, personality, and social interactions~\cite{zhou}. Therefore, recommender systems extended the traditional methods like Collaborative \cite{cf} and Content-based Filtering \cite{cb} to include users' information extracted from their OSNs. In this way, Social Recommender Systems make more personalized suggestions based on an improved user preferences model \cite{Guy:2009:PRS:1639714.1639725}. Several relevant works related to the present paper are discussed next.

\subsection{OSN Analysis to Discover User's Interests}
It has been shown that friends are able to make suggestions in a different number of domains and also share some similar interests \cite{Bonhard:2006:KMK:1188052.1188082}. Therefore, recommender systems might make suggestions for the target user based on her/his friends' preferences. Thus, \textit{social recommender systems} have emerged with the aim of modeling the user's preferences by using the information s/he and their friends have published in OSNs. For instance, the study done in \cite{SinhaS01} demonstrated that friends of the target user provided more useful and better recommendations than recommender systems. Ma \textit{et al.} \cite{Ma:2011:RSS:1935826.1935877} also modeled the preferences of the user in a social recommender system. They took into account that some of the user's friends might have different interests. The premise is that people tend to look for their friends recommendations; hence, this work establishes the difference between trust relationships and social friendships. The authors represent the diversity of tastes among the user's social connections using matrix factorization to improve the accuracy of the recommendations.

In our paper we also consider the exploration of users' connections in the Social Web. However, our approach differs from \cite{SinhaS01} and \cite{Ma:2011:RSS:1935826.1935877}, since the item recommendation for the user may be not only based on his/her direct friends, but also on a community to which the user belongs and which is related to a topic of interest.

\subsection{Social Entity Recommendation on Twitter}

\sloppy
There are two important concerns about information stream personalization (Twitter activity feeds): (\textit{i}) items or news feed filtering of what is to be considered of interest, and (\textit{ii}) relevant content discovery that comes from friends of friends \cite{Chen:2010:STE:1753326.1753503}. In \cite{Chen:2012:CPT:2348283.2348372}, the authors present a framework that merges a traditional collaborative ranking approach with Twitter features such as content information and social relations data, so the model can generate better personalized tweet recommendations. In \cite{shlomo}, the authors make a proposal to solve the news feed filtering problem in OSNs by presenting a method that automatically reorganizes the feeds and filters out irrelevant posts.
The authors in \cite{Hannon:2010:RTU:1864708.1864746} propose a ``users to follow" recommender, implemented by using real time data from Twitter. The details about profiling algorithms and recommending strategies used in their recommender system are presented in http://twittomender.ucd.ie. Each user is modeled considering their recent Twitter activity and their social graph.

Other social entities to recommend to Twitter users are hashtags. Users can add some words prefixed by the symbol \textit{\#} to their tweets and they are identified as hashtags. The hashtags give some relevant meaning and structure to the users' posts as a folksonomy. In \cite{Kywe:2012:RHT:2437747.2437772}, a method that recommends hashtags is presented. It is based on finding similar user-tweet pairs to the target user-tweet pair, so the hashtags used by the neighbors may be recommended.

Compared to the state of the art, our approach may also be used to generate recommendations of news feeds, users to follow, hashtags, and other social entities. However, the novelty of our method is to employ a trending topic of interest to a set of users; consequently, the recommendations that can be generated are topic-dependent and are different for users who are content creators and for those who are distributors.

\subsection{Social Influence and Grouping}

In general, people do not make decisions in a completely rational way; instead they are usually influenced by many factors \cite{Bonhard:2006:KMK:1188052.1188082}. Marketing and e-commerce have exploited data in social network sites to propagate knowledge about products faster and collect users' opinions about them. Depending on these connections, consumer groups or communities are then detected. Dholakia \textit{et al.} \cite{Dholakia2004241} present a model that structures the role of social influence by the community on its members to define its effect at the moment a user makes a choice, participates in collaboration activities, adopts certain behavior or goes into an engagement process. In the model, they set decision making as a direct function of social influence and as an indirect function of worth judgment.

In \cite{Weng:2010:TFT:1718487.1718520}, the study shows the identification of influential tweeters based on their social and commercial importance. The authors propose a method in which the influential users are classified and ranked by topic of interest, and every topic has a small set of representative words associated with it. In \cite{cha2010measuring}, the researchers analyze three measures of influence in Twitter: indegree, retweets, and mentions per user in their dataset, as well as how influence varies across topics. They found that the most influential accounts were authoritative news sources and content trackers (topic independent results).

Some researchers in the field of group recommender systems (which suggest items for a group of friends, a family or a team) have seen that social factors, inherent in human behaviour, influence the recommendation and adoption phases. In \cite{Chen20082082,Quijano,Ye:2012:ESI:2348283.2348373}, the authors study social influence inside groups to evaluate how this can be used to improve group recommender systems design. The work in \cite{Watts441} explains the \textit{two-step flow model of influence} \cite{sociol} where it is said that a small number of people act as influential individuals transmitting information with their own view of mass media to the rest of society. The first step refers to the transmission from the mass media to a group of influential people, and the second step comprises the diffusion of information from the influential group to a bigger audience. Those are the two steps in which a group of leaders may accelerate or prevent an item adoption. From this comes the motivation of our current work to identify influential groups involved in a specific domain of interest in an OSN, where those groups are formed by joining content creators and detecting their corresponding set of distributors. 

The result may be used to build or improve users' preference models and then formulate social item recommendation. This social model has not been proposed before in the related state of the art.

\section{Approach}\label{approach}
This section provides the details of our approach, named $TreToC$ (which stands for ``{\em Tre}nding {\em To}pic {\em C}ommunities''), able to identify content creators and content distributors, as well as detect topic dependent communities related to a trending topic. The approach works in three steps:

\begin{enumerate}
    \item {\bf Identification of $CCs$.} Analyzing the activity of the users who tweeted about a given trending topic, this step identifies the {\em content creators}, {\em i.e.}, those who generate content that is subsequently retweeted by other users.
    \item {\bf Identification of $CDs$.} Analyzing the activity of the users who tweeted about a given trending topic, this step identifies the {\em content distributors}, {\em i.e.}, those who retweet content generated by the creators.
    \item {\bf Detection of Trending Topic Communities.} Given the sets of users detected in the previous two steps, we first generate a graph $G$ that connects them, and then apply a community detection algorithm to detect communities associated to the considered trending topic.
\end{enumerate}

What follows is a systematic account of how the tasks performed by our approach have been implemented. 

\subsection{Identification of CCs}

Users with a certain number of followers, whose tweets are quickly propagated or retweeted because of their content, and who are experts or somehow represent a specific domain, may be considered creators of significant content.

Given a trending topic $h \in H$, we collect the set of tweets $T_h$ that contain $h$ and consider the set of users $U_h$ associated to these tweets ({\em i.e.}, those that either tweeted or retweeted content in $T_h$). Out of all the collected tweets, let $T'_h$ denote the set of tweets that do not represent retweets ({\em i.e.}, those tweets that contain original content). 

Every tweet $t \in T'_h$ is created by users who promote the content amplification over the social network. However, not all the users who generate content can be seen as topic propagators. Indeed, it is essential that the content is considered as interesting by other users, who retweeted a given tweet $t \in T'_h$ at least once. For this reason, we build a set $\hat{T'_h} \subseteq T'_h$, which contains these tweets:
\begin{equation*}
\hat{T'_h} = \lbrace t \in T'_h: retweets(t) > 0 \rbrace 
\end{equation*}
where $retweets()$ is a function that returns the number of times a given tweet was retweeted by other users.

Given the previously defined set, we designate as $CCs \subseteq U_h$ the collection of {\em content creators}, who favor the content generation.  More formally, the set of content creators is defined as follows:
\begin{equation*}
CCs = \lbrace u \in U_h: \exists t \in \hat{T'_h}\ s.t.\ author(t) = u \rbrace 
\end{equation*}
where $author()$ is a function that returns the author of a given tweet.

\subsection{Identification of CDs}
A user who follows another is probably interested in knowing the content s/he posts, but if the user retweets that content as it is, s/he is showing an agreement with it. Moreover, considering the diffusion of a topic, some particular level of interest arises, since many people retweet the emerging tweets. Therefore, the fact that a user \textit{retweets} the tweets of another user is an important source of information to identify the content distributors of a trending topic.
Consider that every user $u \in CCs$ posts a tweet $t \in \hat{T'_h}$. Let $R_t$ be the set of tweets that represent a retweet of $t$:
\begin{equation*}
R_t = \lbrace t' \in T_h \setminus \hat{T'_h}: rt(t', t)=true \rbrace 
\end{equation*}
where $rt()$ is a function that returns true if a tweet $t'$ is originated by a tweet $t$ ({\em i.e.}, if it is a retweet of $t$).

We define as {\em content distributors} ($CDs$) the set of users who retweet content in $\hat{T'_h}$ and act as propagators. More specifically, the set is defined as follows:
\begin{equation*}
CDs = \lbrace u \in U_h: \exists t' \in \cup_{t \in \hat{T'_h}} R_t \ s.t.\ author(t') = u \rbrace 
\end{equation*}

It is worth highlighting that in our approach replies to a tweet are not considered, as they cannot be treated as forms of agreement. It should also be noted that, unlike the \textit{retweeted} content of users,  their \textit{favorited} content is not shown in their followers' timelines; thus, favorite activity does not promote the spread of a topic and it is not considered as part of our study.

\subsection{Detection of Trending Topic Communities}\label{detection}
Given the set of users who generated topic-dependent content ($CCs$) and those who retweeted this content ($CDs$), the first goal is to find an effective way to link them.
Indeed, in order to allow a rapid spread of information, users should follow each other. Moreover, we have to ensure that an explicit connection between a $CC$ and her/his $CDs$ is present.

In order to detect the communities related to a trending topic $h \in H$, it is first necessary to build a graph $G = (V, E)$ that represents the previously mentioned connections. The set $V$ of vertices is represented as the union of the two sets of users identified in the previous two steps:
\begin{equation*}
V = CCs \cup CDs 
\end{equation*}

In order to build the set $E$ of edges that represent the connections among the users, we consider three types of relationships. The first is the following relationship between two topic-dependent content creators:
\begin{equation*}
F_{C} = \lbrace (u_x, u_y) : follow(u_x,u_y)=true, u_x, u_y \in CCs \rbrace  
\end{equation*}
where $follow()$ is a function that returns true if the first user follows the second. 

The second type of connection we consider is the following relationship between two topic-dependent content distributors:
\begin{equation*}
F_{D} = \lbrace (u_x, u_y) : follow(u_x,u_y)=true, u_x, u_y \in CDs \rbrace  
\end{equation*}

In the third type of connection we link a $CC$ to a $CD$ only if the $CD$ retweeted content generated by the $CC$. Note that we avoid adding in the graph the following relationships between $CCs$ and $CDs$ because, in this context, this kind of link would be too generic and too weak to relate two users. Indeed, even if a user follows another, it cannot be taken for granted that these two users agree on everything.

Saying that a Twitter following relationship does not explicitly show dependency to a given topic may sound arbitrary. However, if a user retweets another but does not follow her, and the following relationship would represent the link between two users, there would be no connection between them (even if, with respect to the trending topic, an important connection between the two users exists). Moreover, our focus is to detect communities in which the consumers get in touch with agreeable content with respect to the considered trending topic. Then, the connection between a $CC$ and a $CD$ is well represented by a retweeting link. More formally, the set can be defined as follows:
\begin{multline*}
Ret = \lbrace (u_x, u_y) : \exists (t',t) \in T_h \ s.t.\ rt(t',t)=true \ \wedge \\ author(t')=u_x\ \wedge\ author(t)=u_y \rbrace      
\end{multline*}

Finally, the set $E$ of edges in the graph is represented as:
\begin{equation*}
E = F_{C} \cup F_{D} \cup Ret  
\end{equation*}

At this point, the Louvain method \cite{1742-5468-2008-10-P10008} is applied to detect topic-dependent communities of interest in the graph $G$. The choice of employing a community detection algorithm was made since it can easily handle networks with millions of nodes in a very short time. This characteristic of the algorithm fits with our need to detect communities that rapidly evolve and are characterized by a temporary nature. Given the evolution of a trending topic over time ({\em e.g.}, the appearance of new users that generate new content related to the trending topic), being able to detect communities in a matter of seconds allows the algorithm to work in a real-time scenario like the one we are considering.

Another interesting feature of the Louvain method is its capability to generate communities at different granularities (the structure returned by the algorithm is a dendrogram). Therefore, if a trending topic is emerging, our approach would be able to consider communities at higher granularities to make sure that each community contains both content creators and distributors, and if a topic has existed for a longer amount of time and more users are participating in it, communities at lower granularities might be considered.

As previously mentioned, this capability of the Louvain algorithm to rapidly detect communities would allow to capture a snapshot of the evolution of a trending topic (\textit{e.g.}, at fixed time intervals, it would be possible to re-run the algorithm). However, since our proposal was conceived to provide effective recommendations to the users (both individuals and groups) there would be no need to recompute the communities too many times, to avoid ``flooding" the users interested in the trending topic with excessive information.

\section{Analytical Framework}\label{framework}

This section presents the analytical framework and gives our results. We first present the analytical strategy and setup (Section~\ref{strategy}), followed by a description of the employed dataset (Section~\ref{datasets}) and metrics (Section~\ref{metrics}). Finally, we present the analytical results (Section~\ref{results}).

\subsection{Analytical Setup and Strategy}\label{strategy}

The environment for this work is based on the Python language. To build and manipulate the graph, as well as to calculate the metrics presented next, we used the \textit{NetworkX} module\footnote{https://networkx.github.io}. However, the clustering coefficient of nodes for directed graphs is not part of the functions. Then, we implemented it following its formal definition. To run the Louvain community detection algorithm and measure the graph modularity we used the \textit{community} module\footnote{http://perso.crans.org/aynaud/communities/api.html}. In order to ensure the repeatability of the analyses, some parameters need to be considered:

\begin{itemize}
\item By construction, the graph is {\em directed} and {\em unweighted};
\item To define the communities the function employed was {\em community.best\_partition()} where the {\em resolution} parameter is set to 1. The resolution in modularity is used to adjust the optimization in partitioning. If this value is bigger than 1 it leads to the merging of two communities that share one or more edges, independently of the communities' features. We did not alter the resolution to avoid bias. Because of the properties of Louvain, the directed graph needs to be transformed into undirected when calling the function\footnote{Note that, even though the communities were detected on the undirected graph, the metrics to evaluate their quality were measured on the original directed graph, as described in Section~\ref{detection}.}.
\end{itemize}

The dataset employed in the analyses is the only one existing in the literature containing trending topics on Twitter and the tweets associated to them, which we enriched with the {\em following} relationships between the users, collected thanks to the Twitter API.

To validate our proposal, five sets of analyses were performed:
\begin{enumerate}
\item {\bf Characterization of the trending topics.} Given a trending topic, we analyze the number of content creators and distributors that characterize it. This will allow us to understand the dynamics that characterize the activity on Twitter, even before communities are detected.
\item {\bf Analysis of the disconnected users.} In this case, we analyze the percentage of disconnected users from the graph (which would not be involved in the community detection\footnote{Community detection algorithms work on the largest connected component of a graph.} and thus would not benefit of the information spreading).
\item {\bf Analysis of the cohesion among the users.} For each community, we evaluate its quality by measuring the cohesion between the users in it, using standard metrics such as modularity, ratio between the number of communities and the number of users, and density.
\item {\bf Analysis of the community structure.} For each community, we analyze its composition, by measuring the ratio of content creators and distributors in it, and their clustering coefficient. This allows us to evaluate the effectiveness of our approach to connect those who generate the content to those who make use of it.
\item {\bf Analysis of the relationships between the users.} On Twitter, there are some kinds of relationships that connect people together. Our assumption was that users in a social network might be connected at a given time because of a common topic of interest. However, these users might be associated to a topic because of previous relationships between them ({\em e.g.}, friendship). In order to validate that our communities are topic-based and do not appear together because of previous relationships, for each set of trending topics that share at least one user in common we analyze the percentage of users who take part in the intersection by measuring the Jaccard index. 
\end{enumerate}

In order to verify the choices made in our approach to consider the three previously presented types of connections in the graph, we compare our proposal with a baseline approach named {\em Retweeting-Based Communities} ($RBC$). In the $RBC$ method, the set of edges in the graph connects two users only if one retweeted the other. It is worth mentioning that a graph built on the \textit{following} relationship only would represent a community detection performed on the original Twitter graph, and this is completely unrelated to the trending topic dependency, so we discarded a baseline that considered only this type of connection.

\subsection{Dataset}\label{datasets}

The analyses were performed on a dataset specifically built to collect information about trending topics on Twitter, which was presented in~\cite{ZubiagaSMF15} and is available online\footnote{http://nlp.uned.es/$\sim$damiano/datasets/TT-classification.html}. The dataset contains 1,036 trending topics, which are associated to 567,452 tweets from 348,757 different users. However, in order to form the graph and detect the trending topic communities, the information about the tweets and the users who posted them is not enough. Indeed, we need to have the following relationship between the content creators, and the following relationship between the content distributors (Section~\ref{detection}). This was collected by querying the Twitter API, for the first 368 topics (due to the limitations imposed by the API on the number of calls that could be made). The final dataset contains 67,607 tweets, which correspond to the content retweeted at least once and the retweets found during collection, 15,918 unique creators, and 36,890 unique distributors. Of these, 673 were found to be creators of one topic and distributors of another. If a creator retweeted a tweet in the same topic, s/he was considered only as creator, in order to keep the topic graph structure proper. In conclusion, the total number of users in our study was 52,135, having 29.24\% of them as creators, 69.46\% as distributors, and 1.3\% acting as both (in different topics).

\subsection{Metrics}\label{metrics}
The method we propose produces a graph for a trending topic being analyzed. The graph is then divided into communities of interest. Both the graph and its communities can be evaluated by using the following  metrics. 

\subsubsection{Ratio of disconnected users}
The {\em ratio of users disconnected from the graph} measures the fraction of users, either content creators or distributors, who are not present in the graph because of the lack of linkage.
Let $\overline{V} \subseteq V$ be the subset of users for which there is no edge $e \in E$ that connects them to the graph $G$. The ratio is calculated as follows:
\begin{equation*}
|\overline{V}| / |V|
\end{equation*}

\subsubsection{Cohesion among the users}
After executing the community detection, every node in the graph is going to be assigned to a community. The {\em modularity} is a value that represents the strength of division of a network into communities. High modularity means the connections between the nodes within communities are dense and the connections between nodes in different communities are sparse. The algorithm returns this metric after the community detection process is finished. Readers can refer to~\cite{1742-5468-2008-10-P10008} for further details.

The {\em ratio between the number of communities and the number of users} allows us to evaluate the ability of an approach to group the individual users into communities. Indeed, higher values represent a low cohesion among the users (they are not added to the same community), while lower values indicate a smaller number of communities and higher cohesion among the users.

The {\em density} is the ratio between the number of edges per node to the number of possible edges.
The density of a directed graph $G = (V, E)$ can be calculated as:
\begin{equation*}
|E| / (|V| * (|V|-1))
\end{equation*}

\subsubsection{Community structure}

Every community is expected to have several content creators in order to have newly generated content that can be spread through the community.  The {\em number of creators per community} quantifies how many creators we can find in a community.

The {\em number of content distributors per community} measures how many distributors we can find in a community.

The last metric we are going to use, the {\em community clustering coefficient}, quantifies the extent to which nodes in the graph tend to cluster together. The clustering coefficient for nodes in a directed graph is defined by:
\begin{equation*}
C_i=|\lbrace e_{jk}:v_j,v_k \in N_i, e_{jk} \in E \rbrace|/k_i(k_i-1)
\end{equation*}
where $k_i$ is the number of neighbors of a vertex $v_i \in G = (V, E)$ and $N_i$ is defined by the neighborhood for the vertex: 
\begin{equation*}
N_i=\lbrace v_j:e_{ij} \in E \vee e_{ji} \in E  \rbrace
\end{equation*}

\subsection{Analytical Results}\label{results}
In the following subsections, we provide a detailed evaluation of our proposal.

\subsubsection{Characterization of the trending topics}\label{characterization}
\begin{figure}
\centering
\includegraphics[scale=0.6]{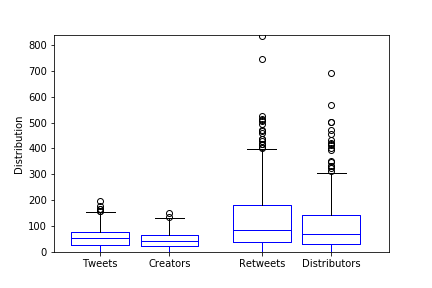}
\caption{Distribution of the number of tweets, creators, retweets and distributors for the Trending Topics.}
\label{fig:Characterization}
\end{figure}

In the following, we analyze what characterizes the trending topics in the dataset. Each boxplot in Figure~\ref{fig:Characterization} represents the number of tweets found per trending topic, the number of creators who posted those tweets, the number of retweets found per trending topic, and the number of distributors who made those retweets. From the distributions obtained as results, we see that content generation and content propagation behave differently. Indeed, the data related to tweets presents a normal distribution (very few outliers) for which the average number of tweets per trending topic is 55.51, while the number of creators per trending topic is 46.20. In contrast, with respect to the content propagation, the distribution is skewed to the right, showing that few trending topics reached a high incidence of retweets/distributors. The median value for the retweets per trending topic is 84.5 and the median number of distributors per trending topic is 69.

\subsubsection{Analysis of the disconnected users}\label{disconnected}
When a user has a connection with another, it is going to be considered in a graph (as a source or destination node, depending on the relationship). Accordingly, the average number of disconnected users for the trending topics was analyzed. Non-linked users are detected once the trending topic graph is obtained by following the corresponding approach, $RBC$ or $TreToC$, while the rest of the users shape the main connected component.

The results show that the trending topic graphs generated with our approach ($TreToC$) cover 85\% of the users who take part of the topic. The $RBC$ baseline loses 23.6\% of the user base of the dataset. These results demonstrate that our graph construction approach (presented in Section~\ref{detection}) includes more users in the community detection process. Indeed, if only the retweets ($RBC$ baseline) are considered, more users are left out of the detected communities with respect to our approach, thus reducing the information spreading.

\subsubsection{Analysis of the cohesion among the users}\label{cohesion}

In order to analyze the level of cohesion between the users in a community, in Table~\ref{tab_cohesion} we report the average values of {\em modularity}, {\em ratio between the number of communities and the number of users}, and {\em density}, for our approach $TreToC$ and the baseline $RBC$.

\begin{figure}
\centering
\includegraphics[scale=0.6]{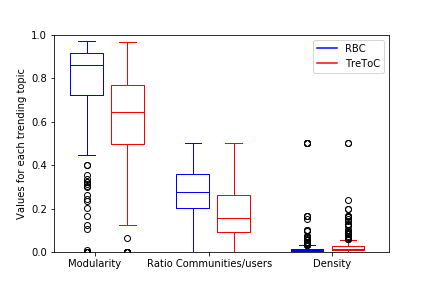}
\caption{Cohesion among the users: Distribution of the metric values.}
\label{fig:metricsGraph}
\end{figure}

The corresponding metric values obtained are presented in Figure~\ref{fig:metricsGraph}. A lower {\em modularity}, as that obtained in the $TreToC$ method, shows that the communities in the graph maintain certain level of interaction or connection between them. This is not seen in $RBC$ graph where the distributors behave only as source nodes, causing the modules partitioning being well defined. In this case, if we would like to adjust the resolution to get fewer communities it would not be possible because the $RBC$ communities are not connected between them. Furthermore, the {\em density} is influenced by this fact, since the users in $TreToC$ share two links ({\em i.e.},  following as well as retweeting) and act as source or destination nodes, resulting in a bigger value compared to density for $RBC$. Note that a higher modularity does not necessarily mean `better', it is better just when we want smaller communities (in terms of number of vertices) or non-connected communities (as the $RBC$ baseline produces). Nevertheless, the purpose of our work is to get fewer communities, which are highly associated units composed by a suitable number of content creators and distributors. For example, more linked content creators in a community would cause diversity in future recommendations.

The average number of communities found in $RBC$ graphs is 26.05, that exceeds the average amount of communities found in the $TreToC$ graphs (16.22 communities per trending topic) which is what our method looks for ({\em i.e.}, our approach obtains larger communities).

\begin{table}
\caption{Cohesion among the users (average).}
\label{tab_cohesion}
\begin{center}
\scriptsize
\begin{tabular}{|c|m{1.38cm}|m{1.4cm}|m{1.1cm}|}
\cline{2-4}
\hline
Method & Modularity & Ratio of Communities/Nodes & Density \\
\hline \hline
RBC & \multicolumn{1}{r|}{0.780} & \multicolumn{1}{r|}{0.278} & \multicolumn{1}{r|}{0.021} \\ \cline{1-4}
TreToC & \multicolumn{1}{r|}{0.622} & \multicolumn{1}{r|}{0.183} & \multicolumn{1}{r|}{0.027}  \\ 
\hline 
\end{tabular}
\end{center}
\end{table}

\subsubsection{Analysis of the community structure}\label{structure}
Table~\ref{tab_structure} shows a summary of the analysis of the community structure obtained for the set of trending topics in our study. More specifically, this analysis measures the {\em average percentage of content creators}, the {\em average percentage of content distributors}, and the {\em average clustering coefficient} for a given community.

The $RBC$ method relates two content creators only if a suitable number of distributors retweeted both of them; just then, we are going to be able to find few creators in a community (in average). In the $TreToC$ method, the following relationship joins content creators making it more likely to find them as close neighbors. Consequently, their individual distributors come together too. We can observe this in the percentage of content distributors in a community in $TreToC$ method, which is bigger too.

As a consequence of being able to have more content creators and distributors linked together in a $TreToC$ community, the {\em clustering coefficient} increases as well compared to the $RBC$ graph.

The boxplots in Figure~\ref{fig:Boxplots, communities structure}, report the results of the three metrics found over the trending topics and compare the two approaches. From the results, we notice that the $TreToC$ method creates proper communities where we can find groups of content creators and the corresponding distributors groups. Notice that to represent the average clustering coefficient in the same figure, the values were multiplied by 100. As the clustering coefficients obtained for the $RBC$ communities had a value of zero, they are not plotted in the figure.

\begin{table}
\caption{Community structure (average).}
\label{tab_structure}
\begin{center}
\scriptsize
\begin{tabular}{|c|m{1.38cm}|m{1.38cm}|m{1.4cm}|}
\cline{2-4}
\hline
Method & \% of Creators & \% of Distributors & Clustering Coefficient \\
\hline \hline
RBC & \multicolumn{1}{r|}{4.54} & \multicolumn{1}{r|}{6.76} & \multicolumn{1}{r|}{0.000} \\ \cline{1-4} 
TreToC & \multicolumn{1}{r|}{7.76} & \multicolumn{1}{r|}{9.05} & \multicolumn{1}{r|}{0.077}  \\ 
\hline        
\end{tabular}
\end{center}
\end{table}

\begin{figure}
\centering
\includegraphics[scale=0.59]{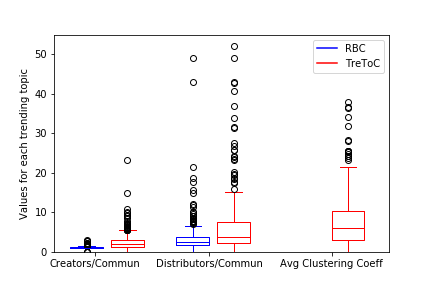}
\caption{Community structure: Distribution of the metric values.}
\label{fig:Boxplots, communities structure}
\end{figure}

\subsubsection{Analysis of the relationship between the users}\label{relationship}
Considering the users who participated in more than one trending topic (3,599 users) and either appear ($i$) as distributors in a given topic and also as creators in other topics (18.7\% of the mentioned 3,599), ($ii$) only as creators in more than one topic (22\%), or ($iii$) only as distributors in more than one topic (59.3\%), we evaluated to which extent those trending topics that share one or more users are overlapped, in order to find out if our communities are topic-dependent or exist because of previous relationships. To do so, we calculated the Jaccard index considering all the users of the set of \textit{possible} overlapped trending topics. We obtained 1,894 different combinations of trending topics that had users in common and the basic statistics show an average Jaccard index of 0.008, having 0.0004 as the minimum value found and 0.65 as the maximum value.

\begin{figure}
\centering
\includegraphics[scale=0.36]{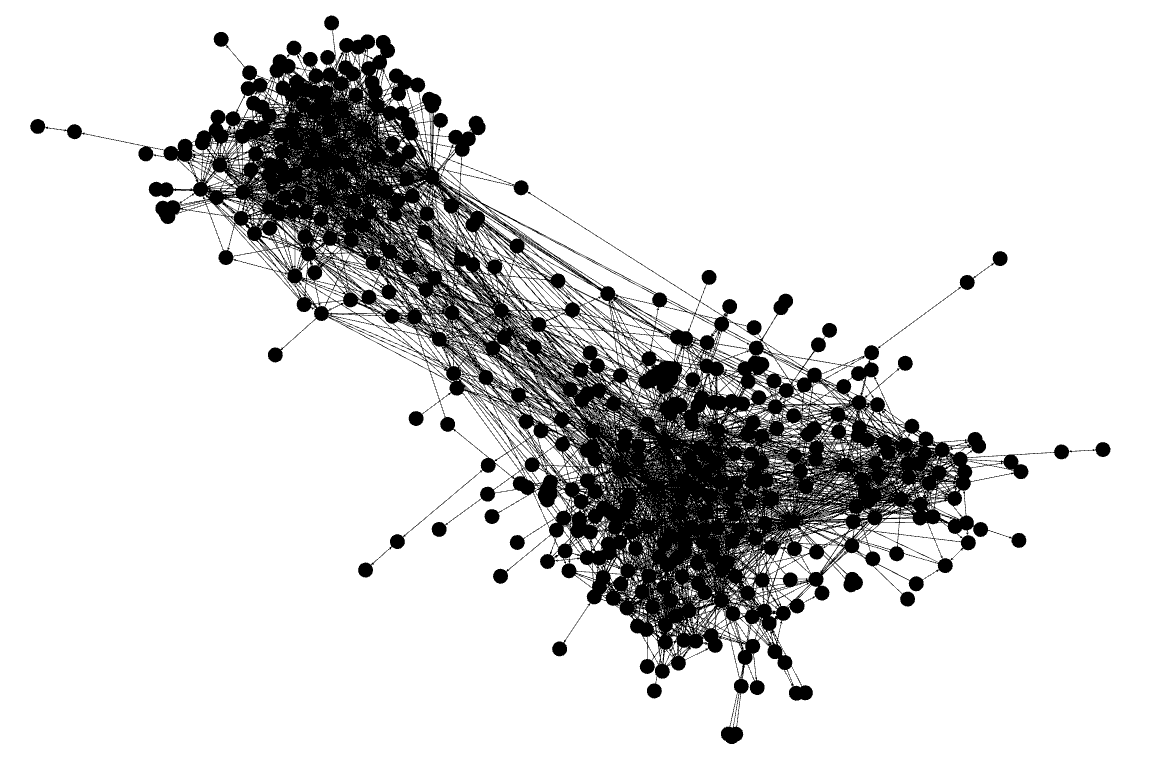}
\caption{Two trending topics graph based on following relationships.}
\label{fig:graph}
\end{figure}

The results validate our approach, whose main focus is to find topic-dependent communities where the users are related to a topic and then linked. The users are gathered together because of the topic and not because of previous relationships between them (indeed, the Jaccard index is very low). As an example, consider the two trending topics `{\tt \#dealwithit}' and `Vernon Gholston', both related to sports and sharing users in common. The hashtag {\tt \#dealwithit} was used by fans of the American football team Buckeyes, who posted tweets like `Go Buckeyes! 93-65 \#dealwithit Wisconsin'. On the other hand, the proper name `Vernon Gholston' belongs to an American football player (who played in Buckeyes). Indeed, the two trending topics are connected between them, hence the overlap between the users. The graph obtained by taking the $CCs$ and $CDs$ for both topics and relating them according to the proposed method (Section~\ref{detection}) is shown in Figure~\ref{fig:graph}. However, despite the shared users, the graph presents two separated groups of participants that are actually dependent of their respective topic, being the {\tt \#dealwithit} group the smallest one.

\subsection{Discussion}

We now summarize the results obtained in our analysis. When working with trending topics, Section~\ref{characterization} showed us that while content generation keeps stable (normal distribution) content propagation is maximized for few trending topics that reach more retweets and distributors than the average. As the analysis of the disconnected users showed (Section~\ref{disconnected}), in order to detect communities that are related to a trending topic and involve most of the users, it is necessary to link the users both with the ``following'' and ``who-retweeted-who'' relationships. Indeed, the retweeting relationship alone leaves around 24\% of the users out of the graph, while the other around 15\%. The analysis of the cohesion among the users (Section~\ref{cohesion}) showed that the communities we created are large (the number of communities is very low if compared to the number of users), that the users in a community are well connected (density is high) and that the communities themselves are connected (modularity is not high); this means that the evolution of a trending topic over time would allow a user to be moved from one community to another, to better fit with her/his current interests and the evolution of the trending topic itself. The third analysis, which studied the structure of the communities (Section~\ref{structure}) showed us that each community contains both a proper number of content creators and distributors (this would allow the distributors to get in touch with diverse content, generated by their content creators counterpart); moreover, the clustering coefficient confirmed that the nodes in the communities tend to cluster well together (the values are high), thus enabling the desired spread of information. The last analysis showed that, even though some topics are related and share users in common, they do not overlap, because the users participating in a given topic depend on it ({\em i.e.}, the communities formed around a trending topic are topic-dependent and do not exist because of other types of relationships).

The results summarized above were monitored while the number of analyzed trending topics increased and they maintained the same tendency for the values obtained in the different metrics. From this we can infer that the captured phenomena are generalizable as trending topic behavior in Twitter.

\section{Conclusions and Future Work}\label{conclusions}

In this work, we have proposed a framework to bring together topic-dependent content creator and distributor groups and identify relations between them.\footnote{Our solution including datasets, code, and results are provided in \url{https://github.com/lore10/Detection-of-Trending-Topic-Communities\_Datasets-Code}.} This approach is new with respect to the state of the art, in which there is a lack of study of $N$ group to $N$ group relationships. Our validation showed the effectiveness of our approach at identifying the proper links between the users who participate in the evolution of a trending topic and then to detect suitable communities, which contain both creators and distributors. 

The last stage of information diffusion is given when the content is presented to the \textit{consumers}. They are the end users who have visibility of the trending topic and related content and are those who follow the creators and distributors. However, a \textit{cold start} problem is evident for ``first time" consumers given that we cannot know if a consumer is really interested in a topic until s/he retweets/favorites a post. In order to mitigate this, text mining can be applied over their historical tweets and content in their Twitter lists where they are subscribed. This is proposed for future work. 

In the context of recommendations, we propose for future work to generate suggestions of social items for groups of Twitter users by leveraging information about their corresponding topic-based creator groups. For instance, the \textit{CDs} can be supplied with a recommendation list of new and relevant users to follow, or a set of the latest tweets corresponding to those users and that fit into their topics of interest. We believe that the recommended items will be optimal for a given time frame, that is, an item is recommended at the time the related topic of interest is actually a current interest for a group. However, the recommendation framework must overcome some of the challenges presented by working with Twitter. For example, Twitter provides relevant information about the topics of interest of a user but it is difficult to quantify the user-to-topic tie strength when the user activity on the social network is very dynamic. That is, a topic may be a trend in a given city for no more than one hour and the number of followers and followees may vary every minute. 

In order to implement a group recommender system based on the current work, our next goal will be to understand the correlation between \textit{CCs} and Twitter influential users. Another challenge is to filter robot accounts that are created to propagate certain kinds of information and give them a top position in trending topics.
This could lead to the identification of non-connected distributors as well as isolated content creators. We propose using machine learning techniques to identify these robots and filter them out.